# Response to "Comment on a recent conjectured solution of the three–dimensional Ising model" by Wu et al.

Z.D. Zhang

*Shenyang National Laboratory for Materials Science, Institute of Metal Research and International Centre for Materials Physics, Chinese Academy of Sciences, 72 Wenhua Road, Shenyang, 110016, P.R. China*

This is a Response to a recent Comment [F.Y. Wu et al., Phil. Mag. **88**, 3093 (2008), also arXiv:0811.3876] on the conjectured solution of the three-dimensional (3D) Ising model [Z.D. Zhang, Phil. Mag. **87**, 5309 (2007)]. Several points are made: 1) Conjecture 1, regarding the additional rotation, is understood as performing a transformation for smoothing all the crossings of the knots; 2) The weight factors in Conjecture 2 are interpreted as a novel topologic phase; 3) The conjectured solution and its low- and high-temperature expansions are supported by the mathematical theorems for the analytical behavior of the Ising model. The physics behind the extra dimension is also discussed briefly.

In the preceding paper, Wu et al. [1] comment on the conjectured solution of the three-dimensional (3D) Ising model presented in [2]. The comments that Wu et al make regarding the presentation (length, usage of some words, placement in Appendix) will not be replied to here. Their comments concerning content concentrate on the low- and high-temperature expansions given in [2] and on the different choices of the weight functions $w_y$ and $w_z$. The latter problem needs clarification; the first two objections have been anticipated and are rejected in [2]. Although it is not necessary to repeat here what is already in the original paper, I shall underline several issues with new ideas.

First of all, as is clear from the references quoted in [2], I do not contest the statement that the Ising model has been well-studied for over 80 years, mainly in great contributions of many distinguished scientists, including the authors of [1]. However, present knowledge cannot serve as a standard for judging the conjectured solution, because the 3D case is not yet fully understood. There are two "dark clouds": 1) the divergence of the so-called "exact" low-temperature expansions and the existence of an unphysical singularity; 2) the possibility of the occurrence of a phase transition at infinite temperature ($T = \infty$, $\beta = 1/(k_B T) = 0$) according to the Yang–Lee theorems [3].

It is regrettable that the objections of the authors of [1] are limited to the outcome of the calculations and that they did not comment on the topology-based approach underlying the derivation. The putative solution was deduced using (among other steps) two conjectures, which at the moment cannot be qualified as rigorous. Therefore, the validity of the solution hinges on the validity of the conjectures. The logic of Conjecture 1 is very simple: The topologic problem of the 3D Ising system, which is the origin of the difficulties, can be dealt with by introducing a boundary condition, i.e., an additional rotation matrix $V'_4$, to smooth the crossings of numerous knots hidden in the boundary condition (equation 15) for the matrix $V \equiv V_3 \cdot V_2 \cdot V_1$ [2]. (The equation number in the preceding sentence and those given later in this article refer to equations in [2]). There are two choices for smoothing a given crossing (×), and thus $2^N$ states of a diagram with $N$ crossings [4]. Mathematically, the state summation $\langle K \rangle = \sum_S \langle K | S \rangle \delta^{\|S\|-1}$, producing the bracket polynomial, appears as a generalized partition function defined on the knot diagram and provides a connection between knot theory and physics [4]. Here, <K|S> is the product of vertex weights, ||S|| - the number of loops in the state S. Therefore, the matrix **V** consists of two kinds of contributions: those reflecting the local arrangement of spins and others reflecting the non-local behaviour of the knots. After smoothing, there will be no crossing in the new matrix $V' \equiv V'_4 \cdot V'_3 \cdot V'_2 \cdot V'_1$ [2], which precisely includes the topologic contribution to the partition function, which becomes diagonalizable. The intrinsic non-local behaviour caused by the knots requires by itself the additional rotation matrix as well as the extra dimension to handle the procedure in the much larger Hilbert space, since in 3D the operators of interest generate a much larger Lie algebra [5]. This merely performs a transformation on the Hamiltonian and the wavevectors of the system. Because the well-recognized "correct" high- and low-temperature expansions never take into account the global topologic effect, they cannot be correct at finite temperatures in 3D. The only exception is that the high-temperature expansions in 3D can be correct at/near $\beta = 0$, where the interaction does not exist (or is extremely weak) so that the global effect is negligible.

I recognize that one of the key assumptions, Conjecture 2 concerning the weight factors $w_x$, $w_y$ and $w_z$, was not presented in a logical sequence in [2], mainly because the details were moved to the Appendices in view of length considerations. The weight factors were defined in the range [-1, 1] and, considering symmetry, their roles can be interchanged without altering the eigenvalues (equation 29) or the partition function (equation 49) (see p. 5372). It is possible to generalize the weight factors in the eigenvectors (equation 33) as complex numbers $|w_x| e^{i\phi_x}$, $|w_y| e^{i\phi_y}$, and $|w_z| e^{i\phi_z}$ with phases $\phi_x$, $\phi_y$, and $\phi_z$. However, only the real part of the phase factors appears in the eigenvalues (29), (30), (31), (49), etc. of the system, so that $w_x$, $w_y$ and $w_z$ can be replaced by $|w_x| \mathrm{Re}|e^{i\phi_x}|$, $|w_y| \mathrm{Re}|e^{i\phi_y}|$ and $|w_z| \mathrm{Re}|e^{i\phi_z}|$, respectively. They may be understood as the results of performing a transformation of the eigenvectors of the 3D Ising system to the "quaternion" Hilbert space and subsequently projecting them back to 3D [2]. Various geometrical phase factors, like the Aharonov-Bohm phase and the Berry phase, among others [6], have been discovered in the past decades, which are related to the global topologic behavior of quantum systems. The potential in quantum mechanics was viewed as a connection that relates with phases at different locations [6], which should be true also for the 3D Ising interactions. The present phase factor, which originates from the geometrical behaviour of the 3D Ising system, is novel. This topologic phase is a function of the interactions and temperature, depending sensitively on whether the knots exist or not.

Thus, the value of the weight factors changes at/near $T = \infty$ owing to the change of the geometrical (topologic) structure, while it crosses over from $|w_x| \text{Re}|e^{i\phi_x}| = 1$, $|w_y| \text{Re}|e^{i\phi_y}| = 0$ and $|w_z| \text{Re}|e^{i\phi_z}| = 0$ (their role can interchange, as mentioned, to maintain the 4-fold integral) for 3D to $|w_x| \text{Re}|e^{i\phi_x}| \equiv 1$, $|w_y| \text{Re}|e^{i\phi_y}| \equiv 0$ and $|w_z| \text{Re}|e^{i\phi_z}| \equiv 0$ (to reduce to the 2-fold integral) for 2D. The latter results in a crossover of the critical exponents. The phase factor $e^{i\phi}$ is akin to the one appearing in Feynman's path-integral theory [7], where the transition amplitude between an initial and a final state is the sum over all paths, connecting two points, of the weight factor $e^{\frac{i}{\hbar}S[x]}$, with $S[x]$ the action of the system. Our action here is topologic, which arises from the overall geometry of the path [6], similar to other topologic phases.

One of the criticisms repeatedly voiced in [1] is based on the "fact" that the convergence of the low- and high-temperature series was rigorously proved. It is argued that the expressions (equations (49), (74) and (99)) cannot be the true solution because the weight factors result in a difference between expressions for the high-temperature limit (equation A.12) and the result for more general temperature (equation A.13), for which $w_x = 1$, $w_y = 0$ and $w_z = 0$ was chosen. But attention has never been paid to the possibility of the existence of a phase transition at/near $\beta = 0$ ([2], page 5371). The Lee-Yang theorems [3], which are rigorous and very general, can be suitable for the 3D Ising model. It would not violate other rigorous results [8], if the singular behaviour at $\beta = 0$ served as a necessary condition adding to the convergence of the series. Proving only the radius of convergence of the series is insufficient (especially in 3D). Lebowitz and Penrose [9] proved a theorem for the high-temperature series and distinguished $\beta > 0$ and $\beta = 0$. They stated clearly that, since $\beta = 0$ lies on the boundary of the region E of (β, z) space, there is no general reason to expect a series expansion of $p$ or $n$ in powers of $\beta$ to converge (p. 102 of [9]). A qualitative picture is given in Griffiths' review [10], showing the shape of the region in the *T-H* plane (Fig. 6) where all is analytic, but he started with the condition $\beta > 0$. The inequality (equation 2B.8) (or other similar ones) of [10], which is important for proving rigorous results, is valid only for $\beta > 0$. Actually, if we plotted Griffiths' *T-H* plane as a *β-H* plane, there should be a singularity at $\beta = 0$. Therefore, distinguishing "at/near infinity" and "finite temperature" is reasonable. Sachdev claimed in Figs. 4.3 and 11.2 of his book [11] that the so-called "lattice high-*T*" phase at very high temperatures has non-universal critical behaviour. Though the singularity in the 1D quantum model (mapping to the 2D Ising model) might not be strong enough to give any sort of transition, it is our understanding that the geometrical change in the 2D quantum model (mapping to the 3D Ising model) may introduce a transition at $T = \infty$. Usually, mathematical theorems [8-10] prove analytical behaviours in a very general form of functions based on some assumptions (for instance, the Peierls condition, $\beta > 0$ for Theorem 2.1. in Sinai [8]; sufficiently small $\lambda$ or $\varepsilon$ in Theorem 18.1.2, Corollary 18.1.4, Theorem 18.3.1, Proposition 18.4.2, Theorem 18.5.1, and assumptions P1, P2 and $E \geq c + 5$ in Theorems 20.3.1-2, 20.4.1-2 and small $\beta$ in (20.5.4) in Glimm and Jaffe [8]), which do not guarantee the analytic behaviour of the low- and high-temperature expansions in their well-known expansion basis (for example,. the divergence of the low-temperature series is contradictory to these theorems). From another angle, we could think that the analytic nature of the expansions for the conjectured solution is supported by these mathematical theorems [3,8-10]. In addition, the conjectured solution reduces to Zandvliet et al.'s results of the anisotropic 3D Ising model where two of the three exchange energies are small compared to the third one [12], which agree with Fisher's rigorous formulae in this limit [13].

The necessity of introducing the extra dimension can be understood from another angle. The basic issues are some key points being often overlooked in quantum statistical mechanics. To introduce the concept of thermal equilibrium (strictly speaking, an undefined (or multidefined) concept), our Ising model is made part of a system big enough for statistical concepts to be useful [14]. In a quantum statistics system, besides the average in a quantum state (expectation value), one also averages with respect to the probability distribution of systems in an ensemble [7,14,15], i.e., a whole collection (a large number $N$) of identical Ising models of $m$ rows and $n$ columns and $l$ planes connected together by *infinitely weak* forces which allow the Ising models to exchange energy but that *do not* contribute to the total energy of the system. Namely, a piece of substance is isolated from everything; any part of the substance must be in equilibrium with the rest serving as a heat reservoir that well defines a temperature [15]. But the temperature in statistical mechanics is actually the time in quantum field theory [7], since the Euclidean time interval can be consistently identified with $\beta$. The partition function $Z = \text{Tr } e^{-\beta H}$ can be represented in the Schrödinger picture as $Z = \int dx \langle x | e^{-\beta H} | x \rangle$, which is merely the transition amplitude with the identification t = $-i\beta$. This indicates that the time *t* is hidden in the framework of the statistical mechanics for an equilibrium system. Therefore, one has a clue that the framework of the statistical mechanics for the 3D Ising systems should include the time, being in the (3+1) dimensional Euclidean spacetime. The same should be true for the 2D quantum model as is shown by the well-known mapping [11].

In quantum mechanics, at any instance of time, the wave function $\Psi$ of a truly isolated system can be expressed by a linear superposition of a complete orthonormal set of stationary wave functions $\Phi_n$: $\Psi = \sum_n c_n \Phi_n$ where $c_n$ is a complex number and is generally a function of time [15]. In quantum statistical mechanics, the wave function $\Psi$ depends on both the coordinates of the system under consideration and the coordinates of the external world (an additional dimension is indeed needed). $\Phi_n$ denotes a complete set of orthonormal stationary wave functions of the system, while $c_n$ is interpreted as a wave function of the external world (depending on its coordinates). Thus, the scalar product ($c_n$, $c_m$) of the *n*th and the *m*th wave function of the external world is also a function of time. This means that the average value of a large number of measurements of an operator, instantaneously given its expectation value, depends indeed on the time, although in the laboratory we measure not its instantaneous value but a time average [15]. However, with the postulates of equal a priori probability and random phase, the wave function of the system can be regarded as $\Psi = \sum_n b_n \Phi_n$ with the phases of the complex numbers $b_n$ being random, to take into account the effect of the external world in an average way. It was emphasized that for this reduction to be effectively valid, the system must interact with the external world. Otherwise, the postulate of random phase is false, because the randomness of the phases means no more and no less than the absence of interference of probability amplitude. However, such a circumstance cannot be true for all time though it may be true at an instant [15]. The postulates of quantum statistical mechanics are regarded as working hypotheses whose justification lies in their agreement with experiments [15]. Such a point of view is not entirely satisfactory and a rigorous derivation is lacking (see pp. 184-188 of [15]). So, the immediate questions are how the system interacts with the external world (it may be somehow inconsistent with what we accepted for *infinitely weak* forces) and what the missing part is during employing the postulates. To answer in detail these questions is beyond the scope of this reply. But the discussions above show the necessity of the extra dimension, and also imply the existence of flaws in the Comment.

In summary, admitting that there are some open questions related to the choice of the weight factors, which will need more research, we have argued that the correct reproduction of the high-temperature expansion cannot be a coincidence and the failure in reproducing term by term the low-temperature expansion does not disqualify the new approach to deal with knots by means of an extension into a fourth dimension.

The author appreciates the supports of the National Natural Science Foundation of China (under grant numbers 50831006, 10674139 and 50331030).